\documentclass[onecolumn,preprintnumbers,amsmath,amssymb,superscriptaddress]{revtex4-2}

\usepackage{epsf}
\usepackage{epsf}
\usepackage{graphicx}
\usepackage{sidecap}
\usepackage{array}
\usepackage{amsmath}
\usepackage{amssymb}
\usepackage{gensymb}
\usepackage{dcolumn}
\usepackage{epstopdf}
\usepackage{bm}
\usepackage{amsmath}

\graphicspath{{./}{../figs/}{../psd/}}

\newcommand\etal{\textit{et al.}}
\newcommand\ud{\mathrm{d}}
\newcommand\tc{T_\text{C}}

\newcommand\mub{\mu_\text{B}}
\newcommand\ef{E_\text{F}}

\usepackage{color}

\def \beq {\begin{equation}}
\def \eeq {\end{equation}}
\newcommand{\req}[1]{Eq.~(\ref{#1})}

\begin{document}
	
\title {Unusual magnetic and transport properties in HoMn$_6$Sn$_6$ kagome magnet}

\author{Firoza~Kabir}
\affiliation {Department of Physics, University of Central Florida, Orlando, Florida 32816, USA}
\author{Randall Filippone}
\affiliation {Department of Physics, SUNY Buffalo State, Buffalo, New York 14222, USA }

\author{Gyanendra~Dhakal}
\affiliation {Department of Physics, University of Central Florida, Orlando, Florida 32816, USA}
\author{Y. Lee}
\affiliation {Ames Laboratory, U.S.~Department of Energy, Ames, Iowa 50011, USA}
\author{Narayan Poudel}
\affiliation {Idaho National Laboratory, Idaho Falls, ID 83402, USA}
\author{Jacob Casey}
\affiliation {Department of Physics, SUNY Buffalo State, Buffalo, New York 14222, USA }

\author{Anup Pradhan Sakhya}\affiliation {Department of Physics, University of Central Florida, Orlando, Florida 32816, USA}
\author{Sabin Regmi}\affiliation {Department of Physics, University of Central Florida, Orlando, Florida 32816, USA}
\author{Robert Smith}
\affiliation {Department of Physics, University of Central Florida, Orlando, Florida 32816, USA}
\author{Pietro Manfrinetti}
\affiliation {Department of Chemistry, University of Genova, 16146 Genova, Italy}
\affiliation {Institute SPIN-CNR, 16152 Genova, Italy}

\author {Liqin Ke}
\affiliation {Ames Laboratory, U.S.~Department of Energy, Ames, Iowa 50011}
\author{Krzysztof Gofryk}
\affiliation {Idaho National Laboratory, Idaho Falls, ID 83402, USA}
\author{Madhab~Neupane $^*$} \affiliation {Department of Physics, University of Central Florida, Orlando, Florida 32816, USA}
\author{Arjun K. Pathak $^*$} \affiliation {Department of Physics, SUNY Buffalo State, Buffalo, New York 14222, USA }





\begin{abstract}
With intricate lattice structures, kagome materials are an excellent platform to study various fascinating topological quantum states. In particular, kagome materials, revealing large responses to external stimuli such as pressure or magnetic field, are subject to special investigation. Here, we study the kagome-net HoMn$_6$Sn$_6$ magnet that undergoes paramagnetic to ferrimagnetic transition (below 376 K) and reveals spin-reorientation transition below 200 K. In this compound, we observe the topological Hall effect and substantial contribution of anomalous Hall effect above 100 K. We unveil the pressure effects on magnetic ordering at a low magnetic field from the pressure tunable magnetization measurement. By utilizing high-resolution angle-resolved photoemission spectroscopy, Dirac-like dispersion at the high-symmetry point K is revealed in the vicinity of the Fermi level, which is well supported by the first-principles calculations, suggesting a possible Chern-gapped Dirac cone in this compound. Our investigation will pave the way to understand the magneto-transport and electronic properties of various rare-earth-based kagome magnets.


	
\end{abstract}
\maketitle
\noindent
\noindent

Topological nontrivial magnetic materials have attracted tremendous attention, and kagome magnets are one of them, which reveal various interesting electronic states such as Dirac fermions \cite{N-1, N-4}, intrinsic Chern quantum phases and spin-liquid phases \cite{GD-15, N-6, N-9, N-10, N-11, N-12}.
Among these kagome materials, transition-metal-based kagome magnets \cite{N-1, N-2, N-3, N-4, N-5, N-13, N-14, N-15, N-16, N-17, N-18} have already appeared to be the distinguished candidates for correlated topological states, as they possess both unusual magnetic tunability and large Berry curvature fields.
These materials are also predicted to support intrinsic Chern quantum phases \cite{Me-9,Me-10} due to their extraordinary lattice geometry and broken time-reversal symmetry.
In addition, a number of nontrivial magnetic phases have been observed in the rare-earth-and-transition-metal-based RMn$_6$Sn$_6$ family (R = rare earth) \cite{R-1, R-2, R-3, R-4, R-5, R-6, R-8}.
Recently, TbMn$_6$Sn$_6$, one of the members of this 166 family, has been identified as a Chern magnet where large anomalous Hall effect (AHE) and the quantized Landau fan diagram featuring spin-polarized Dirac dispersion with a large Chern gap have been observed~\cite{N-11}.
In another member of this family, YMn$_6$Sn$_6$, a topological Hall effect has been revealed~\cite{N-18, New-2, New-3} and explained by a new chirality mechanism originated from frustrated interplanar exchange interactions and the induced strong magnetic fluctuations.
Lately, Gao \textit{et al.} have reported anomalous Hall effect in $R$Mn$_6$Sn$_6$ ($R$ = Tb, Dy, Ho) with clean Mn kagome lattice \cite{NewG}.
Dhakal \textit{et al.} have recently studied magnetic, magneto-transport, and angle resolved spectroscopic measurement of ErMn$_6$Sn$_6$ \cite{G.D}.
Thus, each member of the RMn$_6$Sn$_6$ family is catching extensive attention for experimental exploration.

\begin{figure*}
	\includegraphics[width=18 cm]{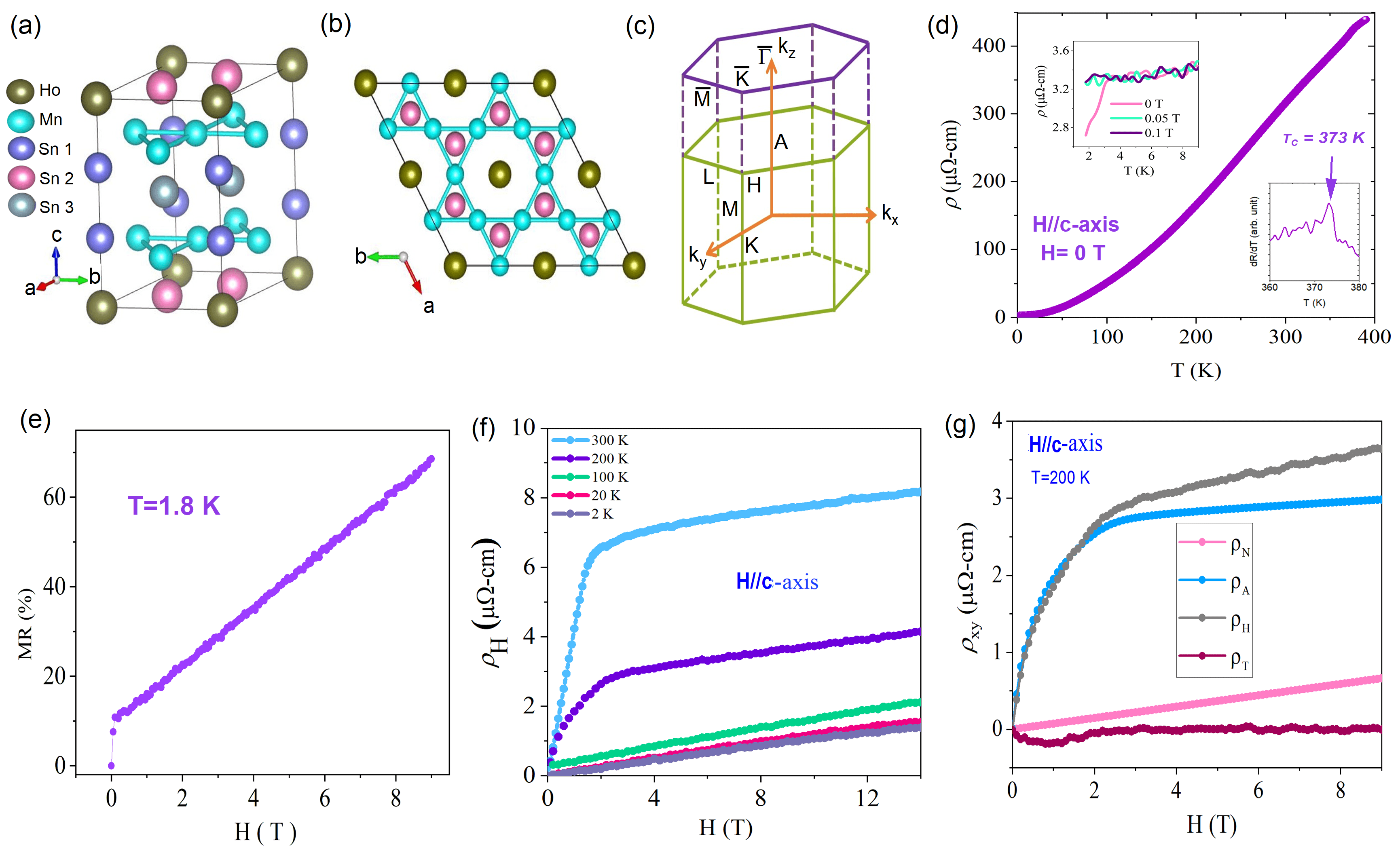}
	\caption{{\textbf{Crystal structure and sample characterization of HoMn$_6$Sn$_6$}}: (a) Crystal structure of HoMn$_6$Sn$_6$. (b) Top view of crystal structure of HoMn$_6$Sn$_6$ forming the kagome lattice. (c) Three dimensional (3D) bulk Brillouin zone (BZ) of the crystal with its projection on the [001] surface. High symmetry points are marked on the plot. (d) Temperature variation of the electrical resistivity of HoMn$_6$Sn$_6$ in zero external magnetic field. The upper inset to the left shows the resistivity at low temperature and magnetic field, while the lower inset to the right indicates the ferrimagnetic transition at \textit{T}$_C$=373 K. (e) Magnetoresistance versus magnetic field measured at $T=$1.8 K. (f) Hall resistivity of HoMn$_6$Sn$_6$ measured at different temperatures with $H\parallel c$. (g) Magnetic field dependence of total Hall resistivity together with the three different components, $\rho^{A}$ , $\rho^{N}$, and $\rho^{T}$ (see text for more details), measured with $H\parallel c$, and at $T$ = 200 K.}
\end{figure*}

In this article, we explore another member of the RMn$_6$Sn$_6$ family, HoMn$_6$Sn$_6$, which consists of two magnetic sublattices: Ho and Mn.
Neutron diffraction studies and magnetic measurements~\cite{Me-30, Me-31, Me-32, New-1} suggested that Mn and Ho sublattices simultaneously transfer from paramagnetic state to ordered states below a critical temperature of $T_\text{C}$= 376~K. With further lowering of temperature, the easy direction of the ordered state starts to reorient from the basal plane toward the $c$ axis below 200 K \cite{Me-31}.
The canting angle with respect to [001] is $48\degree$ at 100 K and remains constant down to 2 K \cite{Me-31}.
The mechanism of the spin-reorientation (SR) transition of HoMn$_6$Sn$_6$ has been quantitatively investigated in a molecular field theory \cite{Me-30}. However, detailed magnetic, transport, and angle-resolved spectroscopic measurements along with theoretical investigation of HoMn$_6$Sn$_6$ have not been reported yet.


Hence, we have performed systematic studies of transport and magnetic behaviors (at ambient and applied pressure) of HoMn$_6$Sn$_6$ kagome magnet. By utilizing angle-resolved photoemission spectroscopy (ARPES), we have measured the electronic structure of this compound, which has been supported by first-principles calculations.
The electrical transport measurement indicates that the compound is metallic and shows a large anomalous Hall effect and topological Hall effect contribution in HoMn$_6$Sn$_6$ at temperatures close to the SR transition temperature, $T_\text{SR}$.
Furthermore, we measure the magnetizations with external fields applied along the in- and out-of-plane directions when the temperature  decreases from 400 K to 2 K.
At $T=2$ K, a well-defined magnetization loop for the field applied along the \textit{c}-axis suggests a strong out-of-plane magnetization component.
Thus, this compound could be a potential candidate for Chern-gapped topological material, according to the Haldane model \cite{Yin9, Yin10, Yin13, Me-9, Me-10}.
Moreover, magnetization measurements under pressure reveal the pressure effects on magnetic ordering at low magnetic fields and below $T_\text{SR}$ (200 K).
ARPES measurements demonstrate the presence of Dirac-like states at the high symmetry point K, close to the Fermi level $E_\text{F}$.
The experimental findings are well supported by first-principles calculations and suggest a possible Chern-gapped Dirac cone in this compound.
Our exploration of electronic and magnetic properties of HoMn$_6$Sn$_6$ will provide an effective way to reveal the magneto-transport behaviors of various rare-earth kagome magnets.

\begin{figure*}[!ht]
  	\centering
  	\includegraphics[width=16 cm]{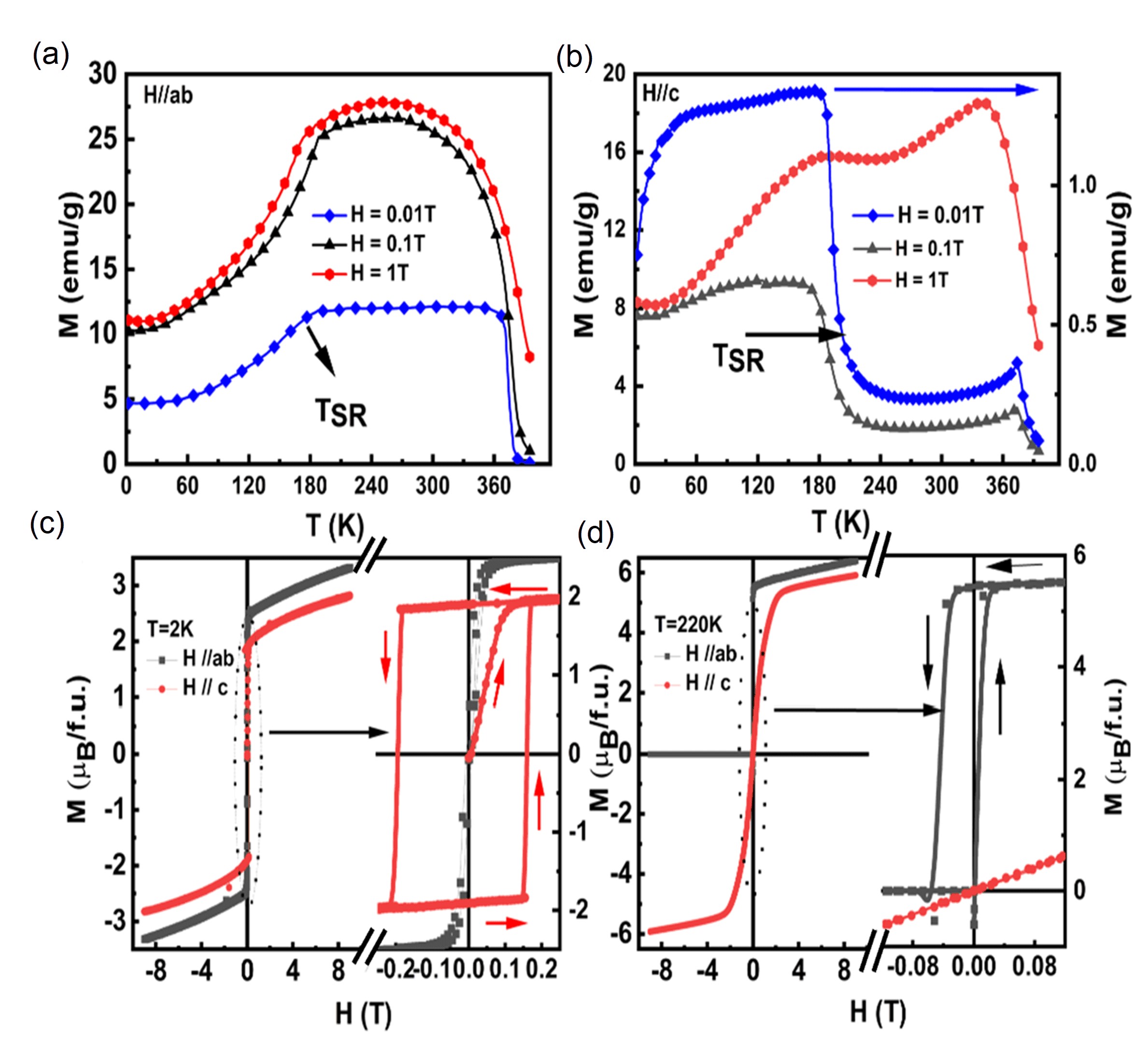}
  	
\caption{\textbf{ Magnetic properties of HoMn$_6$Sn$_6$ single crystals measured along $H \parallel ab$ and $H\parallel c$ (in and out of plane, respectively) and band structures calculated with SOC:} Magnetization as a function of temperature, $M(T)$, measured along (a) $H \parallel ab$ and (b) $H \parallel c$, respectively, at various magnetic fields. Magnetization as a function of magnetic field, $M(H)$, measured along  $H \parallel ab$ and  $H \parallel c$ at (c) $T=$ 2 K and (d) 220 K, respectively. Magnetization at low magnetic field of $M(H)$ for both directions are also shown in the right sides of figures (c) and (d), respectively, for the clarity.}
\end{figure*}

Single crystals of HoMn$_6$Sn$_6$ were grown by the Sn flux technique as described in the supplementary materials (SF.
1).
DC magnetization and transport measurements were carried out using the physical property measurement system (PPMS).
Resistivity and Hall resistivity measurements were performed using the conventional 4-probe method.
A large residual resistivity ratio, RRR = 68, indicates a high quality of the single crystals used in this study.
The electronic structure of HoMn$_6$Sn$_6$ was measured by ARPES at the SLS SIS-X09LA BL at 20 K and under ultra-high vacuum conditions ($5\times10^{-11}$ Torr).
The samples were cleaved \textit{in situ}.
For the synchrotron measurements, the energy resolution was better than 20 meV and the angular resolution was set to be finer than $0.2\degree$.


DFT calculations are performed using a full-potential linear augmented plane wave (FP-LAPW) method, as implemented in \textsc{Wien2K} ~\cite{wien2k}.
The generalized gradient approximation of Perdew, Burke, and Ernzerhof ~\cite{perdew1996prl} is used for the correlation and exchange potentials.
To generate the self-consistent potential and charge, we employed $R_\text{MT}K_\text{max}$ = 8.0 with muffin-tin (MT) radii $R_\text{MT}$ = 2.7, 2.4 and 2.5 a.u., for Ho, Mn, and Sn, respectively.
The calculations are performed with 264 $k$-points in the irreducible Brillouin zone (BZ) and iterated until the total energy difference is lower than 0.01 mRy.
Spin-orbit coupling (SOC) is included with the second-variational method.
Besides using the DFT+$U$ method, we also treated the $4f$ states in an open-core approach to keep the band structures near Fermi surfaces from the influence of the $4f$ states.
The primitive cell contains one formula unit (f.u.).
We adopt experimental lattice parameters~\cite{Me-32} in calculations. 

We construct the TB Hamiltonian by using 118 maximally-localized Wannier functions (MLWFs)~\cite{Pizzi2020}, 59 for each spin channel, corresponding to $d$-type orbitals for Ho and Mn atoms, and $s$- and $p$-type orbitals for Sn atoms in the unit cell.
SOC mixes the spin-up and spin-down states.
The resulting real-space Hamiltonians $H(\mathbf{R})$ with a dimension of $118\times 118$ accurately represent the band structures within the energy window of interest around $E_\text{F}$.

HoMn$_6$Sn$_6$ crystallizes into a HfFe$_6$Ge$_6$-type structure (space group is P6/mmm, (No. 191) ), as shown in Fig. 1(a) with Ho at 1(a) (0, 0, 0), Mn at 6(i) (1/2, 0, z$\sim$0.249), Sn at 2(c) (1/3, 2/3, 0), 2(d) (1/3, 2/3, 1/2) and 2(e) (0, 0, z$\sim$0.34) \cite{Gd_26}, which is composed of Ho layer consisting of Sn atoms and Mn kagome nets, stacked in the sequence -Mn-Ho-Mn-Mn-Ho-Mn- along the $c$-axis \cite{Me-30}.
The Ho and Sn$_2$ atoms lie in the same plane and Mn-Sn$_1$-Sn$_3$-Sn$_1$-Mn atoms are stacked along the $c$-axis alternatively.
The Mn atoms form two kagome layers and the Sn$_2$ and Sn$_3$ atoms form a hexagonal structure.
The hexagonal structure formed by Sn$_2$ atoms can be clearly observed from Fig.~1(b), while viewing along the $c$-axis.
The Ho atoms lie at the centre of the hexagons formed by the Sn$_2$ atoms, clearly visible from Fig. 1(b). Since Sn$_3$ is below the Sn$_2$ layers, hence Sn$_3$ layer is invisible from this top view in Fig, 1(b). The Ho atom lies at the centre of the hexagons surrounded by the Sn$_2$ atoms. On the other hand, being positioned below or above the Mn layers, Sn$_1$ atoms could not be detected from this top view of Fig. 1(b).
Figure~1(c) shows the three-dimensional (3D) BZ and its hexagonal shaped projection onto the (001) plane.


The temperature-dependent resistivity, as plotted in Fig.~1(d), shows the metallic behavior of the sample over the measured temperature range of 1.8 K to 400 K. 
The spin reorientation transition at $\sim$ 200 K is not visible in $\rho(T)$ in Fig.~1(d). However, the ferromagnetic transition of Mn sublattice at $T_\text{C}$=373 K has been observed, clearly seen in $\ud R/\ud T(T)$ (see the inset at the right corner of the Fig .~1(d)). 
The inset at upper left corner shows an anomaly at low temperature ($T=$ 3 K) resistivity at zero magnetic field.
The origin of the anomaly is unclear, but it can be suppressed by the application of a magnetic field as weak as 0.05 T.
Figure~1(e) presents the magnetoresistance ($MR(H)$) of HoMn$_6$Sn$_6$, showing a relatively large value of \textit{MR}=70\% at $\Delta H$= 9 T.
Interestingly, the MR at 2K shows a linear behavior as a function of the magnetic field, suggesting the materials hosting Dirac-like states.
Figure~1(f) presents the magnetic-field dependence of the Hall resistivity in HoMn$_6$Sn$_6$.
As can be seen from the figure, there is a strong contribution of anomalous Hall effect for temperatures above 100 K, and the overall behavior of Hall resistivity resembles the $M(H)$ data shown in Fig.~2(d).

In topological magnetic materials, the total Hall resistivity ($\rho_{H}$) can be expressed as:
\begin{equation}
  \noindent \rho_{H} = \rho^{A}+ \rho^{N}+\rho^{T}
  \label{eq:1}
\end{equation}
where, $\rho^{A} = R_{S}4\pi M$ is the anomalous Hall resistivity, $\rho^{N}(H)  = R_0H$ is the normal Hall resistivity,   and  $\rho^{T}$ refers to the topological Hall resistivity.
\req{eq:1} can be rewritten as $\rho_H = R_{0}H + R_S 4 \pi M$, in the high field saturation region.
We calculate the slope $R_{0}$ and intercept 4$\pi R_S$ from the linear plot of $\rho_H/M$ versus $H/M$ in the high-magnetic field region.
At that region, the ideal linear behavior of the $\rho_H/M$ versus $H/M$ indicates the anomalous Hall resistivity to be the main component of the total Hall resistance.
The $R_0$ derived for $H\parallel c$ and at $T= 200$ K gives positive $\rho^{N}$ (see Fig.~1(g)).
We derive the topological Hall resistivity by subtracting the normal and anomalous component of the Hall resistivity from the total Hall resistivity.
As can be seen from the figure, its contribution is small compared to anomalous Hall resistivity, but its magnitude ($\sim$ 0.2 $\mu$$\Omega$cm) is comparable to other topological materials \cite{N-160, New-3}.\\


 \begin{figure*}
	\centering
	\includegraphics[width=18.0cm]{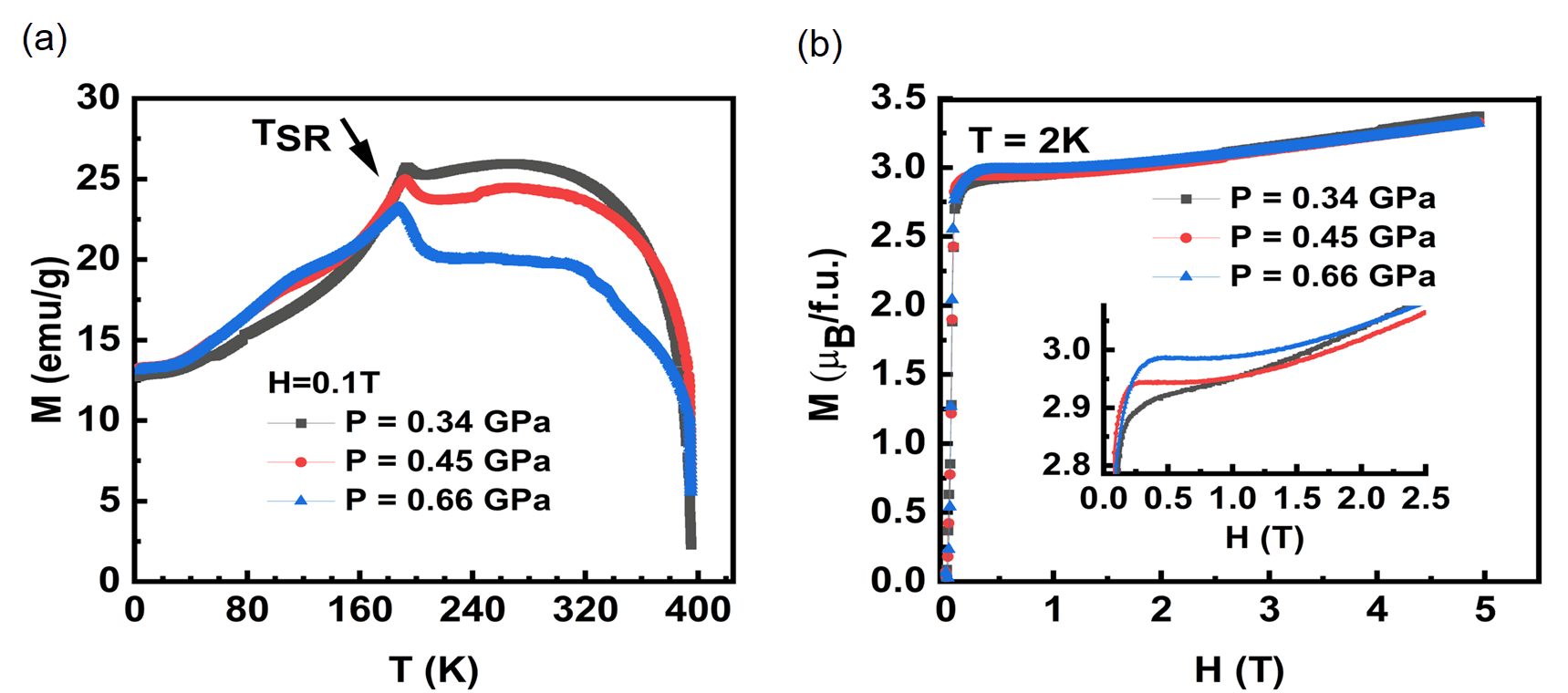}
	\setlength\abovecaptionskip{-15pt}
\caption{{\textbf{Pressure-induced magnetization measurements of HoMn$_6$Sn$_6$ single crystal measured along the $H \parallel c$.} (a) Magnetization as a function of temperature, $M(T)$, measured at various hydrostatic pressure at $H=0.1$ T. (b) Magnetization as a function of magnetic field at various hydrostatic pressure at $T$ = 2K.}}

\end{figure*}
 \begin{figure*}
	\centering
	\includegraphics[width=18.0cm]{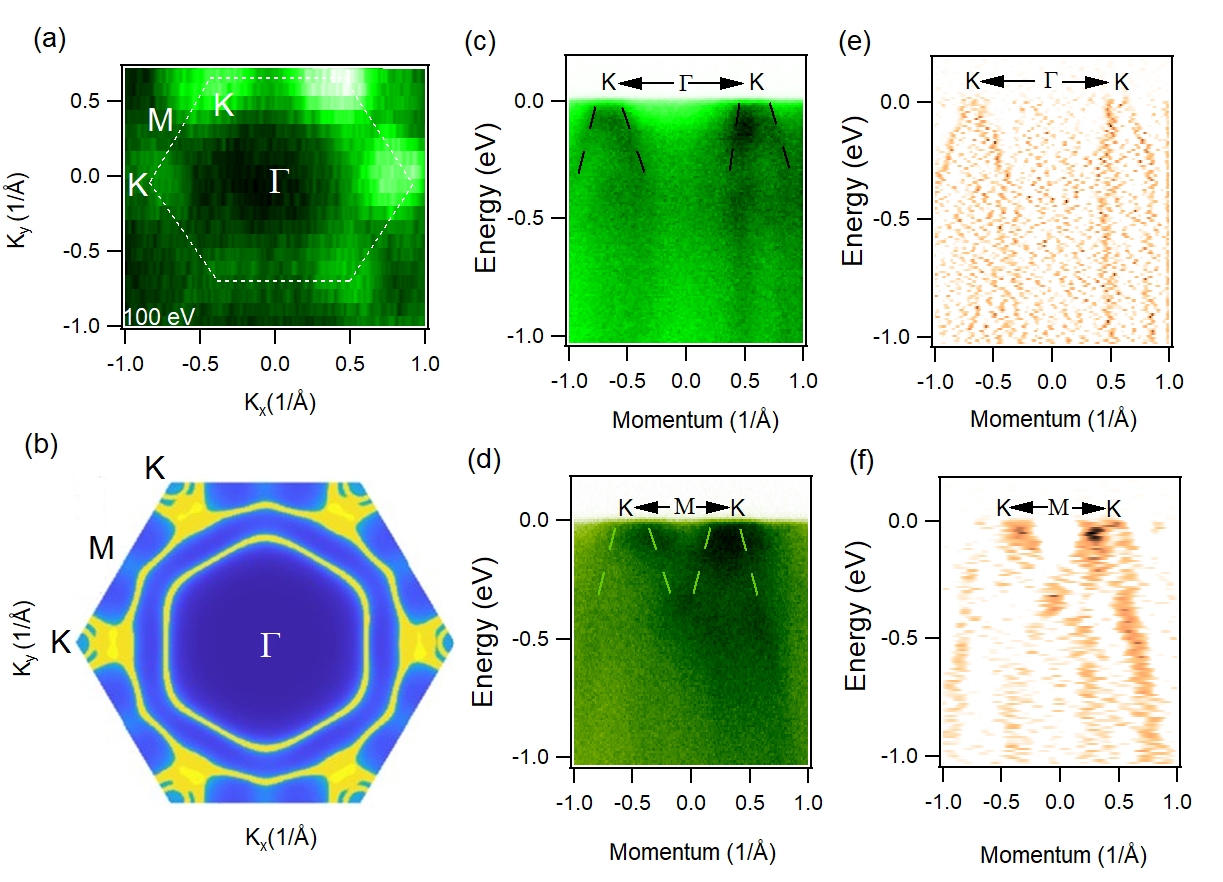}
	\caption{\textbf{Fermi surface and dispersion maps of HoMn$_6$Sn$_6$}. (a) Experimentally measured Fermi surface map. (b) Corresponding calculated Fermi surface map. ARPES measured dispersion maps along the (c) K-$\Gamma$-K  and (d) the K-M-K directions. Second derivative plots along (e) the K-$\Gamma$-K  and (f) the K-M-K directions. All measurements were performed at the SLS SIS - X09LA beamline at a temperature of 20 K with a photon energy of 100 eV.}
\end{figure*}
 
Figures 2(a) and 2(b) show the iso-field magnetization $M(T)$ measured with different in- and out-of-plane magnetic fields, exhibiting two distinct transitions for both directions.
Figure 2(a) presents the $M(T)$ measured at three representative in-plane magnetic fields: $H=$ 0.01, 0.1, and 1 T.
At $H$ = 0.01 T, the magnetization increases smoothly with temperature until $T \leq$ 190 K, becomes plateaued between 200 K to 360 K, finally undergoes transition at 374 K (obtained from minima of $\ud M/\ud T$).
The higher-transition temperature of 374 K at $H$ = 0.01 T is close to the previously reported ferrimagnetic-to-paramagnetic-transition temperature of $\tc$ = 376 K for a polycrystalline sample by Venturini $\etal$~\cite {Me-30}.
The obtained transition temperature at $T=$ 190 K corresponds to the spin-reorientation temperature, $T_\text{SR}$, which compares well with the previously reported value of $T_\text{SR}$= 200 K, obtained from neutron diffraction.
Interestingly, applying external magnetic fields slightly decreases $T_\text{SR}$ but significantly increases $T_\text{C}$.
The ferrimagnetic-to-paramagnetic transition does not complete until 400 K at $H \geq $ 1 T, which is also shown in the non-linear $M(H)$ measured at $T$ = 390 K (supplementary figure, SF.~2).
It is also noticeable that, along $H \parallel ab$ direction, at $H$ = 1 T, the low-temperature ($T \le T_\text{SR}$) magnetization is larger compared to the measurement along $H\parallel c$, and the magnetization difference decreases at $T \ge T_\text{SR}$. The typical isothermal magnetization for both $H \parallel ab$ and $H\parallel c$ at $T=2$ K and 220 K (below and above the $T_\text{SR}$) are shown in Figs.~2(c) and 2(d), respectively. At $T$ = 20 K and $H=$ 9 T, the magnetic moments are 3.14 $\mu_\text{B}$/f.u. and 2.8 $\mu_\text{B}$/f.u. for $H\parallel ab$ and $H \parallel c$, respectively (see SF.~2).
At $T$ = 2 K, the saturation magnetic moments are $M_\text{S}$ = 3.8 and 3.2 $\mu_\text{B}$/f.u.~for $H\parallel ab$ and $H \parallel c$, respectively, obtained from the extrapolation of $M$ vs.~$H^{-1}$ curves.
The value of $M_\text{S}$ is 3.2 $\mu_\text{B}$/f.u., along $H\parallel c$, which is close to 3.26 $\mu_\text{B}$/f.u.,
found by Clatterbuck $\etal$~\cite{Cla}.
The observed saturation magnetic moment values are significantly lower than the saturation moment of Ho$^{3+}$ ion, which is expected for the ferrimagnetic arrangement of Mn and Ho atoms.
The calculated magnetization of the ferrimagnetic configuration is $3.07 \mub$/f.u. in DFT+$U$, agreeing reasonably well with experiments.
The $M(H)$ measured along $H\parallel c$ at $T$ = 2 K, shows a square shaped hysteresis loop with a coercivity of 0.18 T, while the $M(H)$ measured with in-plane fields is non-hysteretic.
The hysteretic behaviors for two field directions reverse above $T_\text{SR}$, suggesting the change of the easy-cone anisotropy at $T < T_\text{SR}$ to the easy-plane anisotropy at  $T> T_\text{SR}$.

The magnetic properties of HoMn$_6$Sn$_6$ were further studied under applied pressure using a Cu-Be mechanical cell with an inner diameter of 2.6 mm, and the lead was used as an internal manometer. The measurements were carried out on the same piece of the sample that has been used for magnetization measurement at ambient pressure (Fig. 2).
The pressure calibration from the manometer is presented in SF. 3. Figure 3 shows magnetization as a function of temperature and magnetic field at pressure up to 0.66 GPa.
The spin-reorientation transition is more pronounced with a peak around 190 K, and the ferromagnetic-to-paramagnetic transition of Mn sublattice is much more sensitive with pressure. The transition temperature shift from 373 K at ambient pressure (Fig. 2(b)) to 394 K at $P$ = 0.34 GPa, with $H$ = 0.1 T (Fig.~3(a)). As shown in Fig.~3(a), it seems that the critical transition temperature is much higher than 400 K at $P > 0.34$ GPa, however, we do not observe that within our measurement range of 400 K.
Considering the value shift of $T_\text{C}$ at $P$ = 0.34 GPa, the $T_\text{C}$ shifts to a higher temperature at the rate of 52 K/GPa, nevertheless, a more detailed $M (T)$ study at pressure and  $T > 400$ K is required.
Fig. 3(b) presents the isothermal magnetization at various hydrostatic pressure at $T=$ 2 K and the inset of Fig. 3(b) shows the pressure dependence of magnetization at low magnetic fields.
The magnetic moment at H = 5 T for both above and below the $T_\text{SR}$ remains relatively same with pressure (Fig. 3(b), and SF. 3(d)).

 


Figure 4 presents the electronic structure of HoMn$_6$Sn$_6$ measured by ARPES and first-principles calculations.
We show the Fermi surface map measured at a temperature of 20 K.
The Fermi map shows the hexagonal symmetry as suggested by the crystal structure.
Furthermore, the Fermi map exhibits the metallic nature, which is accordant with the transport measurements.
Six small circles are observed at the BZ corners K, which display a good hexagonal shape (white dotted hexagon).
These intense circles at the high symmetry points K possibly denote multiple bulk bands and the complex band structure of this material.
However, all features are not visible in the photoemission intensity plots presented in Fig.~4(a).
Figure~4(b) shows the calculated Fermi surface, which is in very good match with Fig.~4(a) and reveals the hexagonal shape of the Fermi map as well. 


Next, we discuss the dispersion maps along the high symmetry directions. Figure~4(c) presents the dispersion map, in which linear Dirac-like dispersive bands exist along the K-$\Gamma$-K path. The 2nd derivative plot of Fig.~4(c) shows the clean Dirac-like dispersion map crossing the Fermi level (Fig~4(e)). Figure 4(d) presents the dispersion map along the K-M-K direction, where electron-like bands exist close to the Fermi level. Corresponding 2nd derivative plot of Fig~4(d) along the K-M-K direction shows some additional bands at M high symmetry point (Fig.~4(f)), where the saddle point of the kagome magnet might exist. Dispersion maps along the K-$\Gamma$-K direction at various photon energies are presented in supplementary section 4 (SF.~4). Besides, the dispersion map and corresponding momentum distribution curve (MDC) and energy distribution curve (EDC) along the K-$\Gamma$-K path are shown in supplementary section 5 (SF.~5). Although HoMn$_6$Sn$_6$ has been suggested to be a chern magnet \cite{N-16}, the possible Chern gap above the Fermi level could not be accessed via our ARPES measurement, therefore, further study is required to support this claim.


\begin{figure*}[b]
	\centering
	\includegraphics[width=18.0cm]{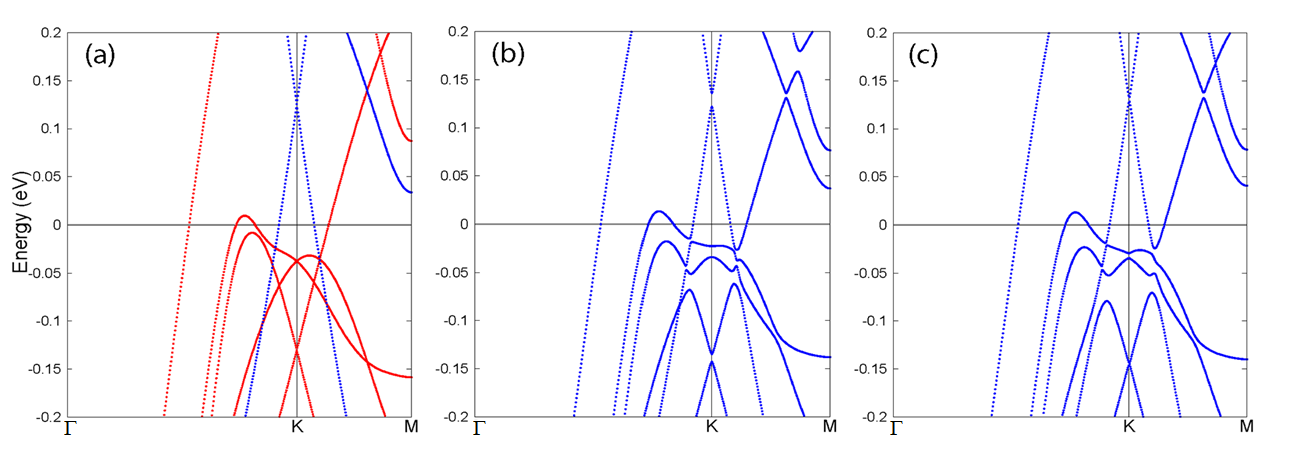}
	\caption{\textbf{Band structures of HoMn$_6$Sn$_6$ along the $\Gamma$-$K$-$M$ path near $\ef$ calculated without and with SOC.} The bands are calculated at $k_z=0.25$ r.l.u.
          (a) Scalar relativistic bands calculated without SOC. (b, c) SOC bands calculated with the spin-quantization direction along the (b) [001] and (c) [100] directions, respectively.}
\end{figure*}

Finally, we discuss the Dirac crossings and SOC-induced gaps and their dependence on the spin orientations.
Figure 5 shows the band structures along the $\Gamma$-K-M path at $k_z=0.25$~r.l.u., calculated with and without SOC.
This particular $k_z$ value is chosen, according to the photon energy used in ARPES, to better compare with experiments. 
As shown in Fig.~5(a), multiple Dirac-like crossings occur near $\ef$ at the BZ corners $K$.
Depending on their band characters, different crossings have different $k_z$ dependence.
For example, the crossing at $\sim$0.12 eV below $\ef$ is much less $k_z$-dependent than the one right above $\ef$, showing a more $2D$-like band character.
Figure 5(b) shows the SOC band structures calculated with the spin-quantization direction along the out-of-plane direction.
As expected, SOC lifts the orbital degeneracy and splits all crossings at K with various gap sizes.
The gaps' sizes depend on the orbital characters of corresponding bands and the SOC Hamiltonian $H_\text{SO}$; the latter depends on the spin direction.
Figure 5(c) shows the SOC band structures calculated with the spin-quantization axis lying in the basal plane.
The SOC-induced gaps become negligibly small or eliminated, in comparison to Fig. 5(b), showing the possibility to control the SOC gap size by changing the spin direction.
Here, we have shown two extreme cases to illustrate the gap evolution with angle change.
Note that, in low temperature, the experimental magnetic ordering direction of HoMn$_{6}$Sn$_{6}$ is tilted to $49\degree$ with respect to [001], resulting in SOC-induced gaps' sizes in between the two cases shown in Figs. 5(b) and 5(c).
Since HoMn$_{6}$Sn$_{6}$ undergoes an SR transition with increasing temperature, the SOC-induced gap should also evolve with temperature.


In summary, we have performed systematic transport, magnetic, and spectroscopic measurements in wide temperature, magnetic field, and pressure ranges of HoMn$_6$Sn$_6$ single crystals. The experimental studies have been supported by first-principles theoretical calculations. Our transport measurement confirms the metallic nature of this compound and shows the signature of a large quantum Hall effect and contribution of topological Hall effect in this system.
From magnetic measurements, we observe a large hysteresis loop and spin reorientation transition below 200 K. We unveil the impact of pressure on magnetic ordering and spin reorientation temperature. Our ARPES measurements reveal the presence of Dirac-like states enclosing the high symmetry point K, supported by the DFT calculations, and suggest the possible existence of a Chern gapped Dirac-like state in this kagome magnet. Altogether, our detailed studies of the kagome magnet, HoMn$_6$Sn$_6$, will provide an ideal platform to understand the magneto-transport properties and the electronic structure of various kagome magnets.\\

%


\noindent \textbf{Acknowledgements}\\~\\
Work at Buffalo State was supported by startup fund from SUNY Buffalo State College and the Office of Undergraduate Research Program. R.F. acknowledges financial support from Office of Undergraduate Research, EURO and Small Grant awards. M.N. is supported by the Center for Thermal Energy Transport under Irradiation, an Energy Frontier Research Center funded by the U.S. DOE, Office of Basic Energy Sciences; the Air Force Office of Scientific Research under Award No. FA9550-17-1-0415, and the Air Force Office of
Scientific Research MURI (FA9550-20-1-0322). Y.~L. and L.~K.~were supported by the U.S.~Department of Energy, Office of Science, Office of Basic Energy Sciences, Materials Sciences and Engineering Division. Ames Laboratory is operated for the U.S.~Department of Energy by Iowa State University under Contract No.~DE-AC02-07CH11358. N.P. and K.G. acknowledge support from the INL Laboratory Directed Research \& Development (LDRD) Program under DOE Idaho Operations Office Contract DE-AC07-05ID14517. We appreciate the support of  Nicholas Clark Plumb, Ming Shi, Hang Li  and Sailong Ju for the beamline assistance at PSI, SLS.\\

*Corresponding author: A.K.P. (Email: pathakak@buffalostate.edu), 
M.N. (Email: madhab.neupane@ucf.edu)\\~\\

\noindent\def\bibsection{

\clearpage

\setcounter{figure}{0}
\renewcommand{\figurename}{\textbf{Supplementary figure}}
\begin{quote}
	\centering
\textbf{Supplementary Materials}
\end{quote}
\bigskip
\bigskip

\noindent
\textbf{Section 1. Crystal growth and energy dispersive X-ray (EDX) spectroscopy of HoMn$_6$Sn$_6$}\\

\noindent The single crystals of HoMn$_6$Sn$_6$ were prepared by the flux growth technique with tin as the flux.  The pure elements in a respective molar ratio of 1:6:30 were loaded in an alumina crucible. Then the crucible was sealed in a quartz tube under vacuum. The quartz tube was heated to 1150$^{\circ}$ C in 8 hrs, dwelled for 10 hrs, and then cooled down to 600$^{\circ}$ C over 140 hrs; finally centrifuged. Large flat hexagonal crystals, as large as 3-5 mm (SF \textbf{1}a), were obtained. To verify the compositional homogeneity of crystals, we performed scanning electron microscopy (SEM) and energy dispersive X-ray (EDX) spectroscopy studies using a Leica Cambridge 360 microscope, equipped with an Oxford X-Max 20 EDX microprobe. The SEM and EDX results confirm the formation of a chemically homogeneous and stoichiometric 1:6:6 phase with no detectable impurities (Table I). \\

\begin{figure*}[b]
\centering
\includegraphics[width=18.5cm]{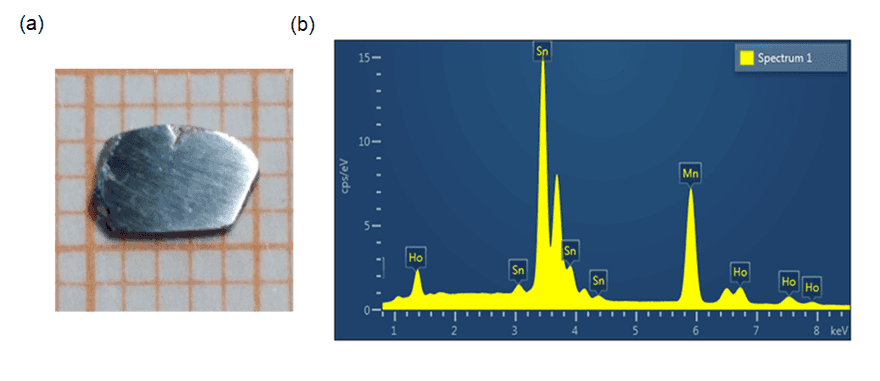}
\caption{\textbf{Crystal growth and energy dispersive X-ray (EDX) spectroscopy of HoMn$_6$Sn$_6$.} (a) Optical image of HoMn$_6$Sn$_6$ single crystal. 
(b) Energy dispersive X-ray (EDX) spectroscopy analysis of HoMn$_6$Sn$_6$.}
\end{figure*}
\newpage
\noindent	
\begin{table}[h]
  \centering
\caption{\textbf{The EDS results showing a high-quality single-phase with a stoichiometry ratio $\sim$ 1:6:6. }}  
\includegraphics[width=18.5cm]{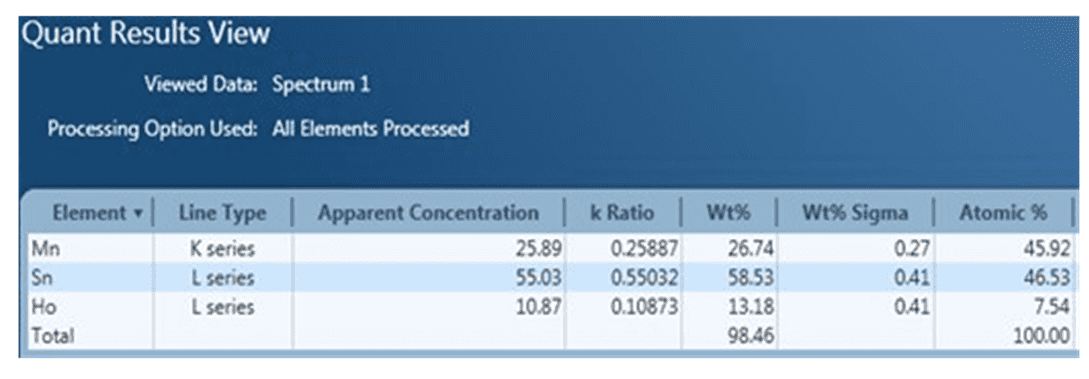}
\end{table}

\newpage 
\noindent \textbf{Section 2. Magneto-transport measurements }\\~\\
Magnetic measurements of HoMn$_6$Sn$_6$ single crystals were carried out for both $H \parallel ab$ and $H \parallel c$ by Physical Property Measurement System (PPMS, Quantum Design), in the temperature range of 2 to 400 K and magnetic field up to 9 T (SF 2). Electrical and Hall resistivity were measured by a standard four-probe technique with an electrical transport option available in PPMS.\\~\\
\noindent
\begin{figure*}[b]
	\centering
	\includegraphics[width=18cm]{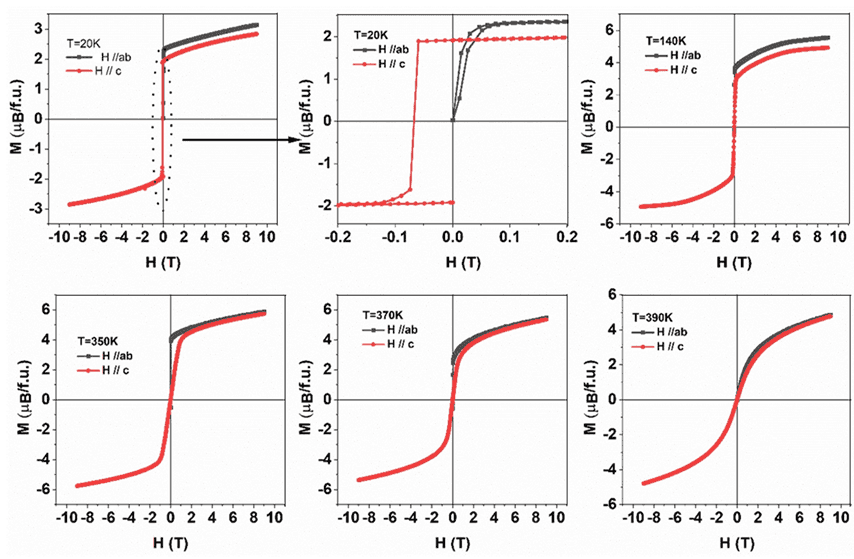}
	\caption{\textbf{Isothermal magnetization curves at various  temperatures ranging from 20 K to 390 K.}}
\end{figure*}
\newpage
\noindent
\newpage
\noindent


\noindent	
\begin{figure*}[b]
\centering
\includegraphics[width=18.5cm]{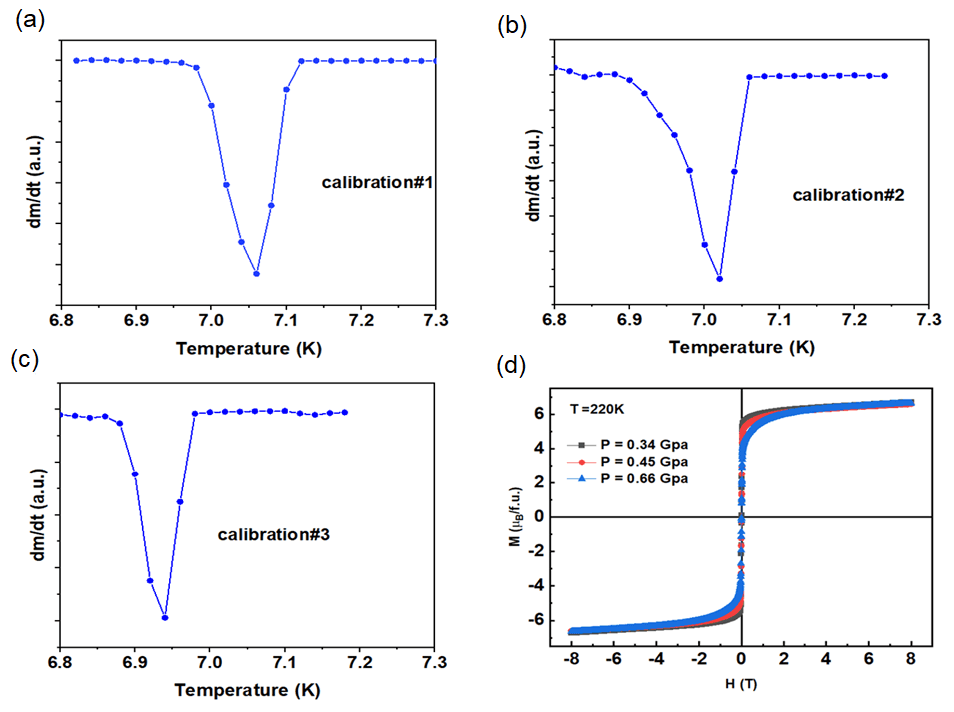}
\caption{\textbf{Pressure determination for magnetic measurements}. (a), (b), (c) The first derivative of magnetization as a function of temperature, dM/dT vs. T, for the lead piece used as a manometer. (d) Isothermal magnetization at $T=$ 220 K at various applied pressure.}
\end{figure*}

\noindent

\noindent \textbf{Section 3. Pressure determination for magnetic measurements}\\~\\
The pressure measurements were carried out at the physical property measurement system by employing a Cu-Be cell manufactured by HMD (type CC-SPr-8.5D-MC4), with lead added as an internal manometer together with the sample, and Daphne (7373) oil used as a pressure-transmitting medium.
A piece of lead wire was used as an internal manometer for the magnetization measurements.
The superconducting temperature of Pb was used to convert into pressure, p = (T-T$_o$)/0.379, where T is the superconducting transition temperature with applied pressure and T$_o$ is the superconducting transition at zero pressure (7.19 K).
SF 3(a), 3(b) and 3(c) show the first derivative of magnetization as a function of temperature, dM/dT vs.~T, for the lead piece used as a manometer.
SF 3(d) presents isothermal magnetization at $T=$ 220 K at various applied pressure.\\~\\

\newpage
\noindent \textbf{Section 4. Photon energy dependent dispersion maps}\\~\\
\noindent
Supplementary figure 4 shows experimentally measured dispersion maps along the K-$\Gamma$-K direction at various photon energies (70 eV, 90 eV and 105 eV), we observe Dirac like feature near the Fermi level at the K-high symmetry points, which suggests the robustness of the Dirac like dispersion at various photon energies.
\noindent
\begin{figure*}[h]
	\centering
	\includegraphics[width=18cm]{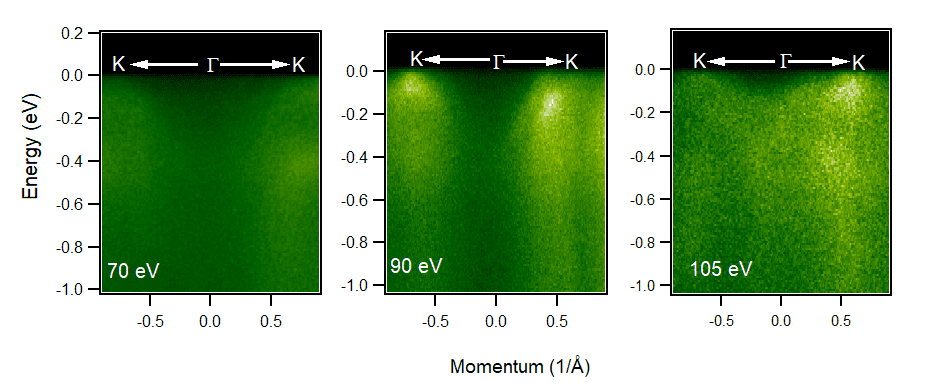}
	\caption{\textbf{Dispersion maps along the K-$\Gamma$-K directions} Dispersion maps at various photon energies, which are noted on the plots. All measurements were performed at the SLS SIS-X09LA beamline at a temperature of 20 K.}
\end{figure*} 

%

\newpage
\noindent \textbf{Section 5. Dispersion maps and corresponding MDC and EDC}\\~\\
Supplementary figure 5(a) shows the dispersion map along the K-$\Gamma$-K direction for photon energy 90 eV, where we observe Dirac cone like features at K high symmetry points close to the Fermi level. Furthermore, MDC and EDC provide the crucial information regarding the bands along the high-symmetry direction K-$\Gamma$-K in supplementary fig. 5(b) and 5(c). Here, we clearly observe the Dirac-like dispersion at the K points, which has been visualized in the MDC peak in supplementary fig. 5(b).\\ 


\noindent
\begin{figure*}[h]
	\centering
	\includegraphics[width=18cm]{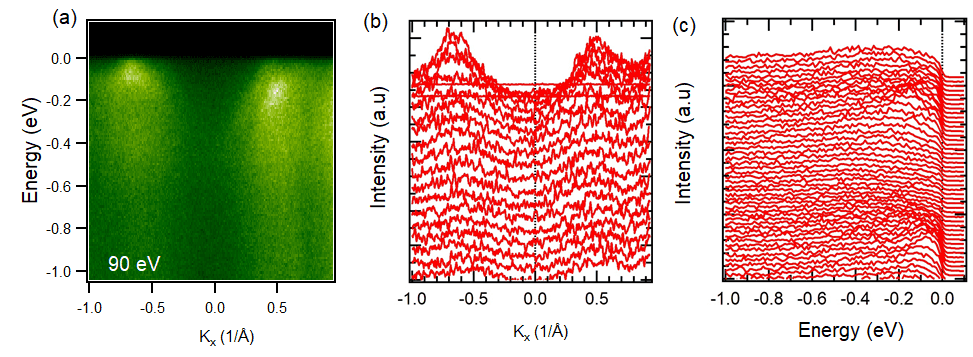}
	\caption{\textbf{Dispersion map and corresponding MDC and EDC along the K-$\Gamma$-K direction} (a) Dispersion maps at photon energy 90 eV along K-$\Gamma$-K direction. (b) Corresponding MDC (c) Corresponding EDC. All measurements were performed at the SLS SIS-X09LA beamline at a temperature of 20 K.}
\end{figure*} 
\noindent

\end{document}